\documentclass[prx,twocolumn,amsmath,amssymb,floatfix,superscriptaddress]{revtex4-1}
\usepackage{graphicx} 
\usepackage{booktabs}
\usepackage{color}
\usepackage[version=4]{mhchem}
\usepackage{xspace, bm, braket}
\usepackage{siunitx}
\usepackage{verbatim}

\usepackage[colorlinks, citecolor=blue, linkcolor=blue, urlcolor=blue, breaklinks]{hyperref}

\newcommand{\SRO}{\ce{Sr2RuO4}}

\newcommand{\eul}{\text{e}}

\def\beq{\begin{equation}}
\def\eeq{\end{equation}}

\begin{document}

\title{Imaginary-time matrix product state impurity solver in a real material
	calculation: Spin-orbit coupling in \SRO{}}

\author{Nils-Oliver Linden}
\affiliation{Arnold Sommerfeld Center of Theoretical Physics, Department of Physics, University of Munich, Theresienstrasse 37, 80333 Munich, Germany}
\author{Manuel Zingl}
\affiliation{Center for Computational Quantum Physics, Flatiron Institute, 162 5th Avenue, New York, NY 10010, USA}
\author{Claudius Hubig}
\affiliation{Max-Planck-Institute for Quantum Optics, Hans-Kopfermann-Strasse 1, 85748 Garching, Germany}
\author{Olivier Parcollet}
\affiliation{Center for Computational Quantum Physics, Flatiron Institute, 162 5th Avenue, New York, NY 10010, USA}
\affiliation{Institut de Physique Th\'eorique (IPhT), CEA, CNRS, UMR 3681, 91191 Gif-sur-Yvette, France}
\author{Ulrich Schollw\"ock}
\email[]{schollwoeck@lmu.de}
\affiliation{Arnold Sommerfeld Center of Theoretical Physics, Department of Physics, University of Munich, Theresienstrasse 37, 80333 Munich, Germany}
\affiliation{Munich Center for Quantum Science and Technology (MCQST), Schellingstrasse 4, 80799 Munich, Germany}

\date{\today}

\begin{abstract}
Using an imaginary-time matrix-product state (MPS) based quantum impurity solver we
perform a realistic dynamical mean-field theory (DMFT) calculation combined with density functional theory (DFT) for \SRO{}. 
We take the full Hubbard-Kanamori interactions and spin-orbit coupling (SOC) into account.
The MPS impurity solver works at essentially zero temperature in the presence of SOC, a regime of parameters currently inaccessible to continuous-time quantum Monte Carlo (CTQMC) methods, due to a severe sign problem.
We show that earlier results obtained at high temperature, namely that the diagonal self-energies are nearly unaffected by SOC and that interactions lead to an effective enhancement of the SOC, hold even at low temperature.
We observe that realism makes the numerical solution of the impurity model with MPS
much more demanding in comparison to earlier works on Bethe lattice models, requiring several algorithmic improvements.
\end{abstract}

\maketitle

\SRO{} attracts much interest due to the superconducting ground state~\cite{Maeno1994,Mackenzie2003a,Mackenzie2017}, the Fermi-liquid behavior~\cite{Mackenzie1996b,Maeno1997,Bergemann2003,Stricker2014}, and characteristics of a Hund's metal~\cite{Mravlje2011}.
In this material the spin-orbit coupling (SOC) affects the pairing symmetry~\cite{Haverkort2008,Veenstra2013} and its inclusion in density functional theory (DFT) was found to improve the description of the Fermi surface~\cite{Haverkort2008, Iwasawa2010}. Meanwhile,
the interplay of SOC and electronic correlations has become a current topic of research~\cite{Behrmann2012, Liu2008,Zhang2016,Kim2018,Tamai2018}. Theoretical predictions of an effective enhancement of the SOC due to electronic correlations by as much as a factor of 2~\cite{Liu2008} have been quantified~\cite{Zhang2016,Kim2018} by realistic DFT plus dynamical mean-field theory (DMFT) calculations~\cite{Georges1996,Kotliar2006} and recently confirmed experimentally~\cite{Tamai2018}. The emerging picture is that SOC changes the electronic structure significantly at wave vectors where nearly degenerate states are found in its absence, but at the same time leaves the overall correlation-induced renormalizations approximately unchanged. However, even with current continuous-time quantum Monte Carlo (CTQMC)~\cite{Gull:2011lr} implementations the Fermi-liquid regime ($T\leq\SI{25}{K}$
~\cite{Hussey1998,Maeno1997,Mackenzie2003a}) is inaccessible~\cite{Mravlje2011, Zhang2016, Kim2018,Tamai2018,Sarvestani2018,Strand2019,Zingl2019}.
This is especially true in the presence of SOC, where CTQMC calculations have only
been performed for $T>\SI{200}{K}$~\cite{Zhang2016,Kim2018,Sarvestani2018,fn1}. 
The sign problem prohibits studying the effect of SOC on electronic correlations at and below the Fermi-liquid temperature, yet it is essential for the precise characterization of the Fermi-liquid state 
and the superconducting instability emerging out of it at $T\sim\SI{1.5}{K}$~\cite{Maeno1994,Mackenzie2003a,Mackenzie2017},
making methodological progress desirable~\cite{fn3}.

Alternative multiorbital impurity solvers to gain insights on such regimes are zero-temperature matrix-product state (MPS) based methods~\cite{white92,schollwock05,schollwock11} or more generally tensor-network based approaches~\cite{bauernfeind17}. Pioneering work~\cite{garcia04,karski05,karski08} considered single-orbital DMFT, using traditional frequency-space density-matrix renormalization group (DMRG) methods~\cite{hallberg95,kuhner00,jeckelmann02}. Progress towards multiorbital calculations was made after introducing Chebyshev-based frequency-space methods~\cite{holzner11,wolf15} and time-evolution methods~\cite{vidal04,daley04,white04,verstraete04ripoll} combined with the linear prediction of long-time behavior~\cite{barthel09}. The first real-time (real-frequency) results for two orbitals or two dynamical cluster approximation (DCA) patches~\cite{wolf14,ganahl14,ganahl14i} were in excellent agreement with CTQMC results for Green's functions on the imaginary-frequency axis~\cite{wolf14} and yielded superior results on the real-frequency axis. To overcome bad computational scaling in the number of orbitals or DCA patches, an imaginary-time evolution was used to study a three-orbital Hubbard-Kanamori Hamiltonian on a Bethe lattice~\cite{wolf15iii}. Recently, efficient representations of the impurity problem as a tensor network~\cite{bauernfeind17,bauernfeind18} have been pursued to perform multiorbital DFT+DMFT calculations directly on the real-frequency axis. Only problems which simplify due to diagonal Green's functions and self-energies have been studied in the literature with MPS/tensor-network-based methods so far.

In this work, we use the MPS-based imaginary-time impurity solver~\cite{wolf15iii} in a DFT+DMFT calculation for \SRO{} with realistic band structures and a Hubbard-Kanamori interaction. The Green's functions and self-energies become matrix-valued and off-diagonal due to the presence of SOC. After carefully benchmarking the MPS solver in the case without SOC against CTHYB~\cite{Werner:2006rt, Gull:2011lr, TRIQS/CTHYB} at very low temperature ($\SI{30}{K}$), we study \SRO{} at low temperature in the presence of SOC -- a regime inaccessible with CTQMC~\cite{Zhang2016,Kim2018,Sarvestani2018}.
We confirm that the effective enhancement of the SOC obtained at room temperature in Refs.~\cite{Zhang2016, Kim2018} still holds
even at zero temperature. Remarkably, the complexity of the MPS solver
due to the entanglement is much larger for this kind of realistic calculation than for a Kanamori model on the Bethe lattice with the same number of orbitals, requiring various algorithmic improvements. Our results show the power of this approach and put earlier DFT+DMFT results on a firm footing. 

The low-energy physics of \SRO{} comprises the three \ce{Ru}-$t_{2g}$ orbitals $m \in \{ xy, xz, yz\}$ hybridized with O-$2p$ states. These states can be well described with three maximally localized Wannier functions~\cite{MLWF1,MLWF2} of $t_{2g}$ symmetry centered on the \ce{Ru} sites~\cite{Tamai2018}. Like in multiple recent works~\cite{Tamai2018, Zingl2019, Strand2019}, we use a three-orbital single-particle Hamiltonian $\hat{H}$ resulting from a Wannier function construction based on a non-SOC DFT calculation. The details of the Hamiltonian construction can be found in
Ref.~\cite{Tamai2018}. The SOC is then added as an atomic term, as described below.

We add a Hubbard-Kanamori interaction~\cite{Kanamori1963,Georges2013}
\begin{align}
\hat{H}_{\text{int}} =  (U-3J) \frac{\hat{N}^2-\hat{N}}{2} - 2J\hat{{\bf S}}^2 - \frac{1}{2} J\hat{{\bf L}}^2 + \frac{5}{2} J\hat{N} \,
\label{eq:HwithoutSOC}
\end{align}

\begin{align}
U ~~~ (U - 2J) ~~~~ (U -3J)
\end{align}
with the number of particles $\hat{N}=\sum_{m\sigma} \hat{n}_{m\sigma}$; total spin on a site
$\hat{{\bf S}}=\tfrac{1}{2} \sum_m \sum_{\sigma\sigma'} \hat{d}^\dagger_{m\sigma} \text{\boldmath$\sigma$}_{\sigma\sigma'} \hat{d}_{m\sigma'}$ and orbital isospin
$\hat{L}_m = i \sum_\sigma \sum_{m'm''} \epsilon_{mm'm''} \hat{d}^\dagger_{m'\sigma} \hat{d}_{m''\sigma}$. $\hat{d}^\dagger, \hat{d}$ are fermionic creation/annihilation operators and $\text{\boldmath$\sigma$}$ are Pauli matrices; the indices $m,m',m''$ run over orbitals and $\sigma\sigma'$ over spins. We use $U=\SI{2.3}{eV}$ and $J=\SI{0.4}{eV}$, as in several other works~\cite{Mravlje2011,Stricker2014,Tamai2018,Zingl2019,Strand2019}.

In DMFT, the Hubbard-Kanamori model, defined by $\hat{H}+\hat{H}_{\text{int}}$, is mapped to a multiorbital impurity problem consisting of an interacting site, capturing all the atomic physics, coupled to an effective noninteracting bath. The impurity site is defined by the interactions (\ref{eq:HwithoutSOC}) and the term $\hat{H}_{\text{chem}}=\sum_{m\sigma} (\epsilon_{m\sigma}-\mu)~ \hat{n}_{m\sigma}$, where $\epsilon_{m\sigma}$ are the averaged band energies and $\mu$ is adjusted to enforce an average occupation $\langle \hat{N} \rangle = 4$ for \SRO. The bath is determined in a self-consistent manner.

\begin{figure}[t]
	\includegraphics[width=0.96\linewidth]{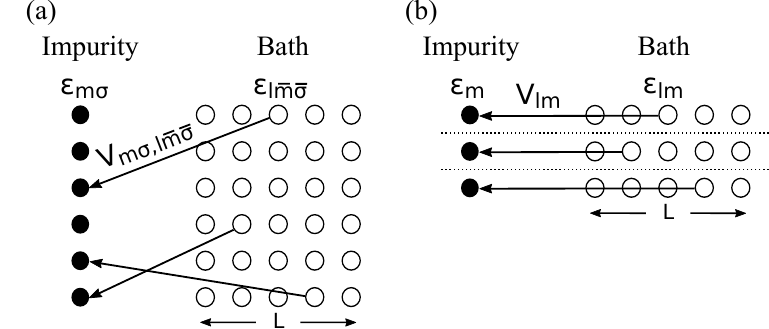}
	\caption{(a) Generic setup of impurity and bath for a three-orbital
	problem: The six orbital and spin degrees of freedom of the impurity are
	labeled $m\sigma$; to each, a bath of length $L$ is attached, with
	labels $l\bar{m}\bar{\sigma}$, where $\bar{m}\bar{\sigma}$ label the
	associated impurity degree of freedom, $l=1,\ldots,L$. The bath is
	characterized by on-site energies $\epsilon_{l\bar{m}\bar{\sigma}}$ and
	hopping elements $V_{m\sigma,l\bar{m}\bar{\sigma}}$ connecting all bath
	degrees of freedom with all impurity degrees of freedom. (b) In the
	case of \SRO{} without SOC, the absence of off-diagonal single-particle
	terms and spin degeneracy reduce the problem to hoppings only between
	each of the three impurity orbitals (grouping the two spin degrees of
	freedom) and their associated baths.}
	\label{fig:figbath}
\end{figure}

Unlike action-based CTQMC algorithms, MPS impurity solvers need a Hamiltonian form of the bath, representing its hybridization with the interacting impurity orbitals using a generalized form of the
Caffarel-Krauth procedure~\cite{caffarel94}. The bath and hybridization Hamiltonians take the form (see Fig.~\ref{fig:figbath}(a))
\begin{equation}
\hat{H}_{\text{bath}} = \sum_{l \bar{m} \bar{\sigma}} \epsilon_{l \bar{m} \bar{\sigma}} \hat{c}^\dagger_{l \bar{m}\bar{\sigma}} \hat{c}_{l \bar{m}\bar{\sigma}} 	
\end{equation}
and
\begin{equation}
\hat{H}_{\text{hyb}} = \sum_{m \sigma,l\bar{m}\bar{\sigma}} V_{m\sigma,l\bar{m}\bar{\sigma}} \hat{d}^\dagger_{m\sigma} \hat{c}_{l \bar{m}\bar{\sigma}} + \text{h.c.} 
\label{eq:hhyb}
\end{equation}

Here, the operators $\hat{c}^\dagger, \hat{c}$ act on the bath. ($m$,$\sigma$)
and ($\bar{m}$,$\bar{\sigma}$) label the orbitals and spin states of the
impurity site and the associated bath sites; these sites we label by $l$ such
that $(l\bar{m}\bar{\sigma})$ uniquely labels a bath operator. The hopping
elements $V_{m\sigma, l\bar{m}\bar{\sigma}}$ are in general complex and connect
{\em any} impurity orbital with {\em any} bath orbital. The bath on-site
energies $\epsilon_{l \bar{m} \bar{\sigma}}$ and hopping elements
$V_{m\sigma,l\bar{m}\bar{\sigma}}$ are determined in the self-consistency cycle
of DMFT, which produces both noninteracting and (using the MPS formalism for
the Hamiltonian
$\hat{H}_{\text{int}}+\hat{H}_{\text{chem}}+\hat{H}_{\text{bath}}+\hat{H}_{\text{hyb}}$)
interacting Green's functions $G^0, G$
with $G^{(0)}_{m\sigma,m'\sigma'}(\tau) \equiv - \langle \mathcal{T}_\tau
\hat{d}_{m\sigma}(\tau) \hat{d}^\dagger_{m'\sigma'}(0)\rangle$ and thereby
hybridization functions $\Delta_{m\sigma,m'\sigma'}(\tau)$~\cite{Georges1996}.
The hybridization function of the discrete bath representation is related to the bath parameters as 
 \begin{equation}
 \label{eq:hyb}
 \Delta^{\text{disc}}_{m\sigma,m'\sigma'}(i\omega_n) = \sum_{l \bar{m} \bar{\sigma}}  \frac{V^*_{m\sigma,l\bar{m}\bar{\sigma}} V_{m'\sigma',l\bar{m}\bar{\sigma}} 
 }{i\omega_n-\epsilon_{l \bar{m} \bar{\sigma}} }\, ,
 \end{equation}
 with Matsubara frequencies $i\omega_n \equiv i(2n+1)\pi/\beta$. 
 Operating in frequency space, we minimize the cost function
 \begin{equation}
    C = \sum_n \omega_n^{-1} \| \mathbf{\Delta} (i\omega_n) - \mathbf{\Delta}^{\text{disk}} (i\omega_n) \| .  \label{eq:matrixfitting}
 \end{equation}
 Note that at $T=0$ the hybridization function (\ref{eq:hyb}) is defined on the continuum, but for fitting purposes, we evaluate it at discrete points, which for reasons of comparability we choose to be the Matsubara frequencies of the fictitious temperature $\beta_{\text{eff}}=\SI{200}{eV^{-1}}$~\footnote{We perform the fit in the frequency range $\omega_n \in [0,6]$, which corresponds to 190 fitting points}.

\begin{figure}[t]
	\includegraphics[width=1.0\linewidth]{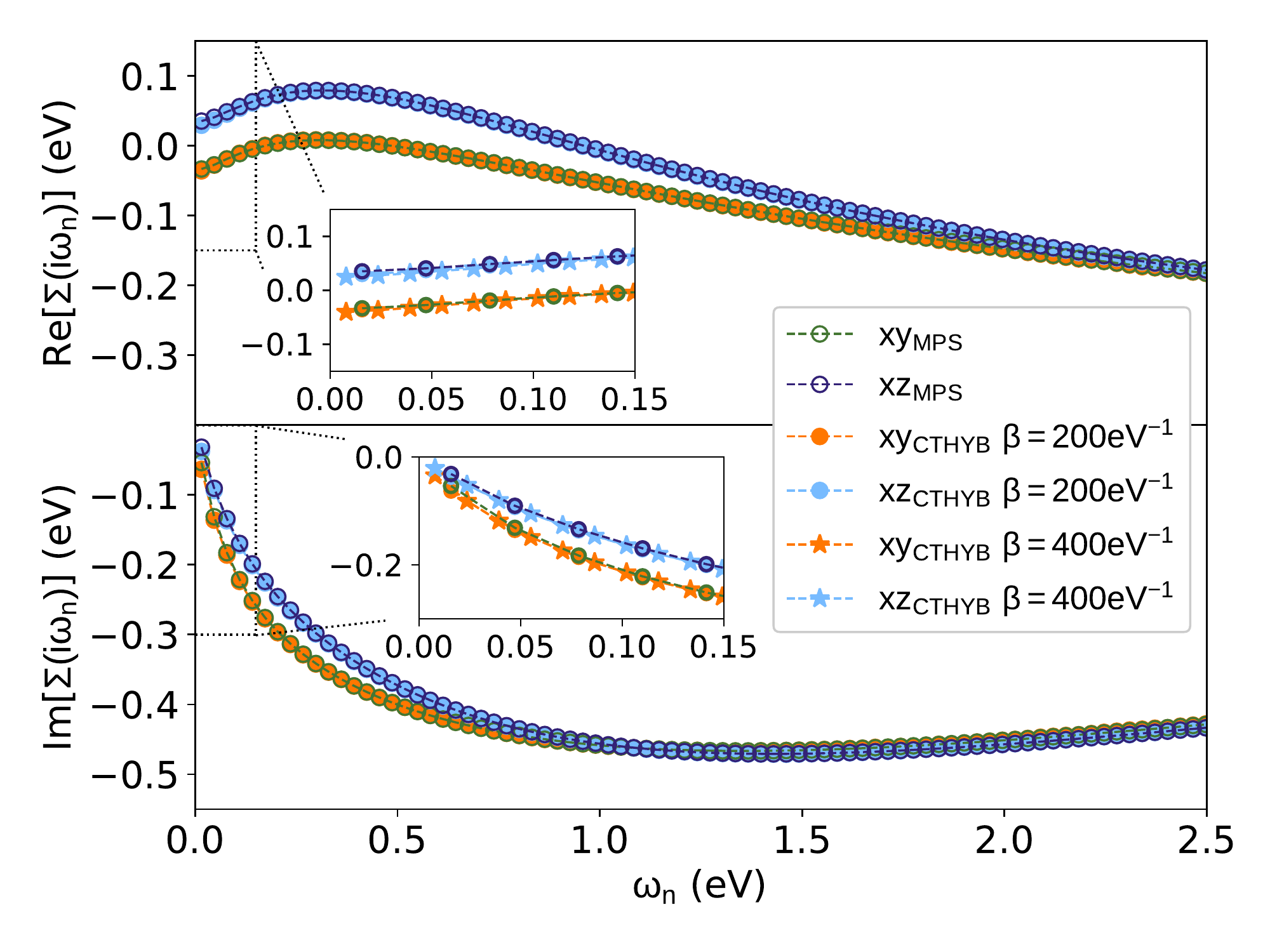}
	\caption{Real (top) and imaginary (bottom) parts of the DFT+DMFT self-energies $\Sigma_{m}(i\omega_n)$ for the $xy$ and $xz/yz$ orbitals without SOC using the MPS impurity solver and a bath size of $L=8$  (open circles). For \SRO{} without SOC the self-energy matrix is diagonal in the orbital basis and the $xz/yz$ orbitals are degenerate. Results are compared to self-energies obtained with CTHYB as impurity solver at $\beta = \SI{200}{eV^{-1}}$ (full circles) and $\beta = \SI{400}{eV^{-1}}$ (full stars), i.e., 58 and \SI{29}{K}. The chemical potential has been subtracted from the real parts.}
	\label{fig:fig2b}
\end{figure}

In general, the fitting problem is challenging, because the Green's functions, the self-energies, and the hybridization functions of the impurity site are matrix-valued (here $(6\times 6)$ for three orbitals with two spin states each).
However, drastic simplifications occur in the case of \SRO{} without SOC:
Due to the symmetries of the crystal structure the Green's functions, self-energies, and hybridization functions are diagonal matrices in the orbital/spin basis. This means we can restrict each bath to couple only to the corresponding orbital and spin state,   
\begin{equation}
\hat{H}^{\text{noSOC}}_{\text{hyb}} = \sum_{l m\sigma} V_{lm\sigma} \hat{d}^\dagger_{m\sigma} \hat{c}_{lm\sigma} + \text{h.c.} 
\end{equation}
In the absence of a field, there is no spin dependency, simplifying the hopping elements further ($V_{lm\sigma}\rightarrow V_{lm}$) and leading to the hybridization $\Delta_{m\sigma,m\sigma}\rightarrow\Delta_{m}$ (see Fig.~\ref{fig:figbath}(b)). This is the case that has been explored by multiorbital MPS-based solvers in the literature so far~\cite{wolf14,ganahl14,ganahl14i,wolf15iii,bauernfeind17,bauernfeind18}. Due to the greatly simplified bath structure, we can then obtain the parameters $V_{l m}$ for a three-orbital model from three individual scalar fits of the hybridization function $\Delta_{m}$, as done in Ref.~\cite{caffarel94}. For a bath of length $L$, we therefore have to solve three fitting problems with $2L$ real fitting parameters each. Note that for \SRO{} there are actually only two fitting problems because of the $xz/yz$ degeneracy. After fitting, we determine the ground state of the model $\hat{H}_{\text{int}}+\hat{H}_{\text{chem}}+\hat{H}_{\text{bath}}+\hat{H}_{\text{hyb}}$ within an MPS-based ground-state search~\cite{hubig15}. We determine the Green's function on the Matsubara axis by an imaginary-time evolution, combined with a linear prediction for longer times~\cite{barthel09}, and a subsequent Fourier transformation to close the DMFT self-consistency loop~\cite{Supp}.

\begin{figure}[t]
	\includegraphics[width=1.0\linewidth]{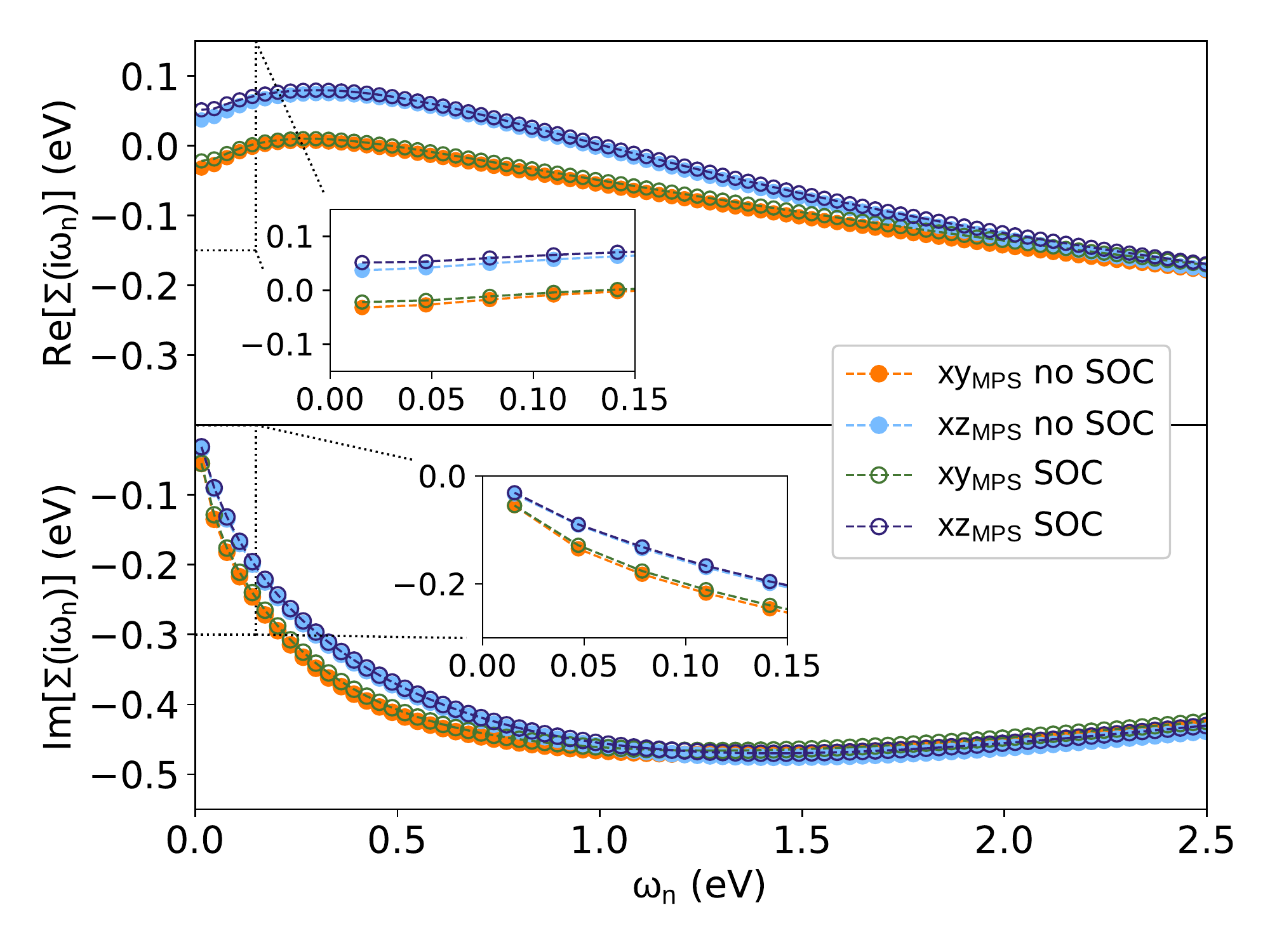}
	\caption{Real (top) and imaginary (bottom) diagonal elements of the DFT+DMFT self-energies $\Sigma_{m\sigma,m'\sigma'}(i\omega_n)$ for the $xy$ and $xz/yz$ orbitals without (full circles) and with (open circles) SOC using the MPS impurity solver and a bath size of L=4. The chemical potential has been subtracted from the real parts.}
	\label{fig:fig3}
\end{figure}

In Fig.~\ref{fig:fig2b}, we show the low-frequency behavior of the self-energy $\Sigma_m(i\omega_n)$, and compare it to CTHYB data obtained at inverse temperatures
of $\beta=\SI{200}{eV^{-1}}$ and $\beta=\SI{400}{eV^{-1}}$~\cite{fn5}.
Overall, the excellent agreement with CTHYB
results fully validates our MPS impurity solver. We attribute the small discrepancies
to the finite temperature, as the results in Fig.~\ref{fig:fig2b} for $L=8$ are already well
converged in the bath size~\cite{Supp}.

Compared to earlier calculations on the Bethe lattice~\cite{wolf15iii}, the computational cost of the impurity solver in this realistic setting increases by about an order of magnitude and numerical convergence into the ground state of the impurity model becomes much slower.
This is the consequence of a strong increase of the ground-state entanglement, which then persists in the time evolution.  We emphasize that the comparison shown in Fig.~\ref{fig:fig2b} is numerically challenging. Multiple algorithmic aspects of the ground-state search and the imaginary-time evolution are crucial to reach such an agreement: both the ground-state search and the time evolution have to be modified from standard procedures to ensure correct results~\cite{Supp}.

We now turn to calculations with SOC included. SOC plays an essential role in the low-energy physics of \SRO{}, as it lifts degeneracies in the bands found in its absence, which especially impacts the shape of the Fermi surface~\cite{Sigrist2002,Eremin2002,Haverkort2008,Iwasawa2010,Puetter2012,Veenstra2013,Scaffidi2014,Steppke2017,Zhang2016,Kim2018,Tamai2018}. Expressed in the cubic $\{ xy, xz, yz\}$ basis, the SOC 
can be approximated as a $k$-independent term, which reads in second-quantized form (cf. Ref.~\cite{Tamai2018})
\beq \label{eq:hsoc}
\hat{H}_{\mathrm{SOC}}\,=\, \frac{\lambda}{2} \sum_{mm'}\sum_{\sigma\sigma'} \hat{d}^\dagger_{m\sigma} \left(\mathbf{l}_{mm'} \cdot \text{\boldmath$\sigma$}_{\sigma\sigma'}\right) \hat{d}_{m^\prime\sigma'}\, ,
\eeq
where $\mathbf{l}$ are the $t_{2g}$-projected angular momentum matrices and $\lambda$ is the coupling constant. In the following we use $\hat{H}+\hat{H}_{\text{SOC}}$ (with $\lambda=\SI{0.11}{eV}$) as the noninteracting part of the Hubbard-Kanamori Hamiltonian. 

Due to the SOC, the Hamiltonian is now complex valued, such that we have to allow for complex hoppings to the bath in DMFT, and, even more importantly, we now must allow for the general ``off-diagonal'' couplings $V_{m\sigma,l\bar{m}\bar{\sigma}}$ and a lifted spin degeneracy due to the form of the SOC term. The matrix-valued fitting problem (\ref{eq:matrixfitting}) simplifies, because an inspection of~(\ref{eq:hsoc}) reveals 
that the orbital/spin states couple in two groups of three each, simplifying the fitting problem to two $(3\times 3)$ problems (for baths of length $L$, we have $3L$ real on-site energies and $3\times 3L$ complex hoppings, i.e., $21L$ real parameters). A further simplification is possible by a transformation from the cubic harmonic basis to the $J$ basis, where the SOC term becomes real and diagonal. The single-particle contribution to the impurity Hamiltonian is diagonal for two of the six states and has the form of two $(2\times 2)$ blocks for the other four (due to the degeneracy structure of \SRO: $\epsilon_{xz} =  \epsilon_{yz}  \neq \epsilon_{xy}$). The total Hamiltonian of the impurity site becomes real and has good quantum numbers $\hat{n}$,$\hat{J}_z$ which we implement. We represent the bath using real parameters, and we face two scalar fitting problems with $2L$ real parameters and two matrix-valued fitting problems with $2L + 4L=6L$ real parameters. For the matrix-valued fittings we improve numerical stability by a two-step procedure: As we expect the off-diagonal elements to be small in comparison to the diagonal elements, we first fit the dominant diagonal terms. Then, we used the obtained parameters as an initial guess for the fits of the $(2\times 2)$-matrix-valued hybridization function. All these considerations are important
for a strong acceleration of the MPS ground-state search and time evolutions. With SOC included we
are limited to $L=4$~\cite{fn6}. Note, however, that in this setup we have a total number of fit parameters being larger than in the $L=8$ case without SOC. Additionally, the $L=4$ self-energy is already well converged in comparison to $L=8$ at low energy without SOC~\cite{Supp}.

\begin{figure}[t]
	\includegraphics[width=1.0\linewidth]{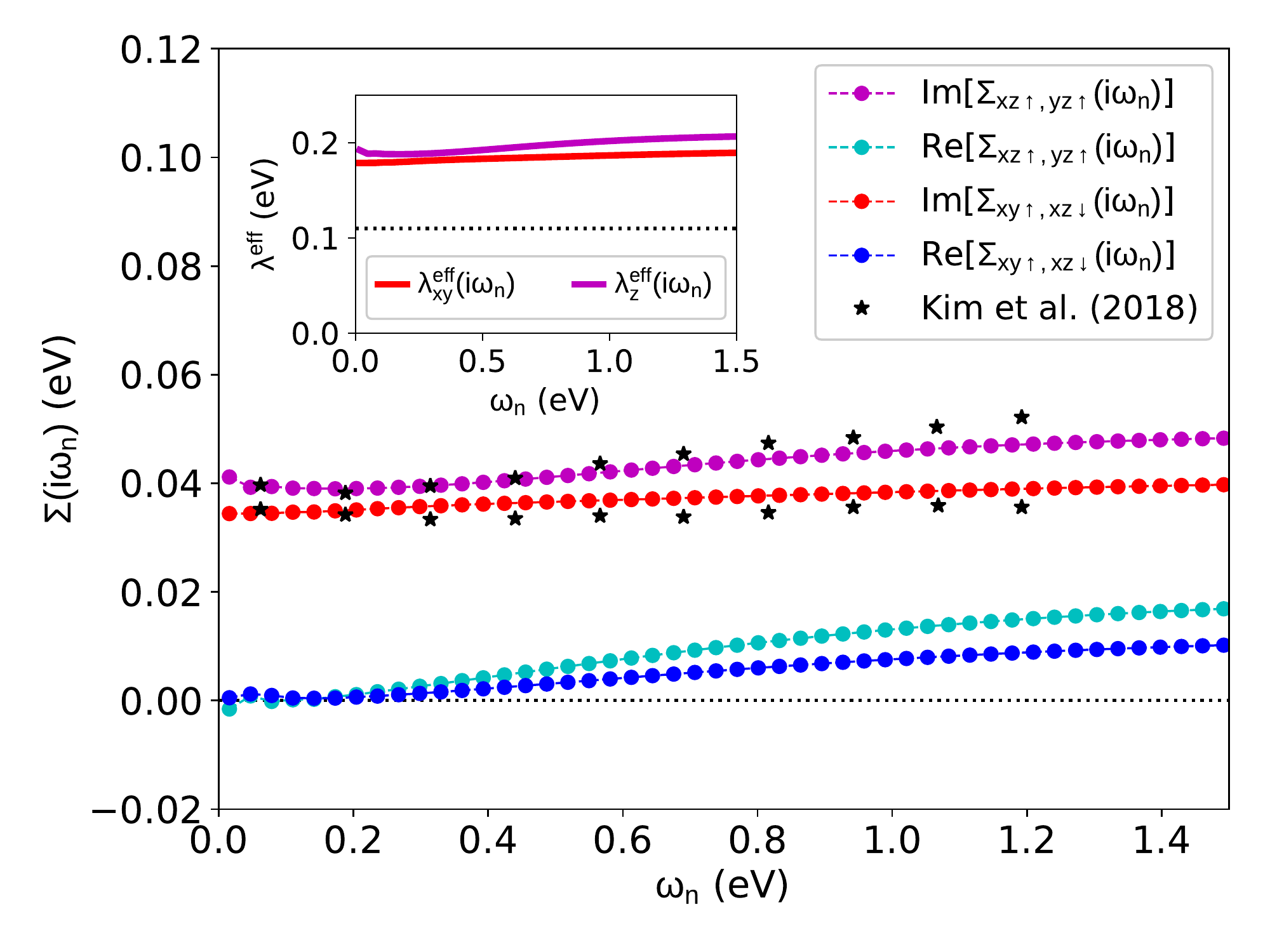}
	\caption{Selected off-diagonal parts of the self-energy $\Sigma_{m\sigma,m'\sigma'}(i\omega_n)$ for a bath size of $L=4$ per orbital and spin state. The excellent agreement with CTHYB results obtained at $T=\SI{232}{K}$~\cite{Kim2018} (black stars) suggest that the self-energy off-diagonal elements are almost temperature independent. For completeness the full self-energy matrix is shown in the Supplemental Material~\cite{Supp}; however, all elements are related to those shown here. The inset shows the effective SOC taking the effect of the off-diagonal self-energy into account.} 
	\label{fig:fig5}
\end{figure}

The diagonal elements of the self-energy, shown in Fig.~\ref{fig:fig3}, are found to be essentially unchanged by the SOC term, which complements earlier CTHYB findings at higher temperature~\cite{fn1,fn2}.
Correspondingly, the quasiparticle renormalizations, $Z_{xy} = 0.19$ and $Z_{xz}=0.29$, are also nearly unaffected by the SOC, $Z^{SOC}_{xy} = 0.20$ and $Z^{SOC}_{xz}=0.27$~\cite{fn4}. The orbital-dependent
mass renormalizations serve as an important benchmark for the MPS impurity solver, as they are well
known from both theory and experiment~\cite{Bergemann2003, Mravlje2011, Zhang2016, Kim2018, Sarvestani2018, Tamai2018}. We note in passing that the strong renormalizations are to a large degree a consequence of the Hund's coupling and the van Hove singularity, where the latter makes the $xy$ orbital more correlated than the $xz/yz$ ones~\cite{Mravlje2011, Kugler2019}.

The off-diagonal self-energies $\Sigma_{m\sigma,m'\sigma'}(i\omega_n)$ are nonzero only for indices coupled by SOC (i.e., $\mathbf{l}_{mm'} \cdot \text{\boldmath$\sigma$}_{\sigma\sigma'} \neq 0$) and vary only slowly with frequency (see Fig.~\ref{fig:fig5}), especially when compared to the frequency dependence of the diagonal elements~\cite{Supp}. This observation is in agreement with previous studies at higher temperature showing that interactions lead to approximately constant off-diagonal terms in the self-energy which can be interpreted as an effective enhancement of the SOC~\cite{Liu2008,Kim2018,Zhang2016,Tamai2018} (see Fig.~\ref{fig:fig5}, black stars). We obtain effective couplings of $\lambda^{\mathrm{eff}}_{z} = \SI{192}{meV}$ and $\lambda^{\mathrm{eff}}_{xy} = \SI{179}{meV}$ (Fig.~\ref{fig:fig5}, inset), in good agreement with Ref.~\cite{Kim2018}, where a simplified two-dimensional tight-binding model was used at $T=\SI{232}{K}$. Note that this characteristic of the off-diagonal elements is not generic, but rather specific to the case of \SRO{}~\cite{Triebl2018}.

In conclusion, we embedded the MPS impurity solver in a DFT+DMFT framework with full Kanamori interactions and spin-orbit coupling. We put our focus on one well-studied material, \SRO, and performed calculations without SOC, where we were able to benchmark the results with precise CTHYB data.
Then we demonstrated the power of MPS by including SOC at zero temperature; a regime inacessible to available Monte Carlo-based impurity solvers. Surprisingly, we found the MPS entanglement to be larger than in similar model calculations on the Bethe lattice, which required some algorithmic improvements.
We confirmed that the main effects of electronic correlations in \SRO{} are an orbital-dependent mass renormalization with $Z_{xy} > Z_{xz/yz}$ and an effective enhancement of the SOC strength by a factor of about 2. The fact that we found an excellent agreement with literature~\cite{Bergemann2003, Mravlje2011, Zhang2016, Kim2018, Sarvestani2018} validates the MPS impurity solver,
but also puts assumptions on the self-energy at low temperatures made previously~\cite{Tamai2018, Zingl2019, Strand2019} on a more solid footing. These works relied on approximating the correlation effect of SOC by a twice as large SOC strength compared to the bare DFT value.

SOC might be also a key ingredient when it comes to magnetic ordering in \SRO{}: Experiments show no magnetic ordering~\cite{Mackenzie2003a}, which is in disagreement with DFT+DMFT calculations without SOC~\cite{Strand2019}. The MPS impurity solver constitutes a viable alternative for realistic DFT+DMFT calculations in regimes inaccessible to Monte Carlo algorithms, and thus we are convinced that it is a method perfectly suited to address such questions in the future.

~\\
We thank Antoine Georges, Andrew Millis and Ara Go for useful discussions. C.H. acknowledges funding through ERC Grant QUENOCOBA, ERC-2016-ADG 423 (Grant No. 742102). N.O.L. and U.S. acknowledge support by the Deutsche Forschungsgemeinschaft (DFG, German Research Foundation) under Germany's Excellence Strategy-426 EXC-2111-390814868 and by Research Unit FOR 1807 under Project No. 207383564. U.S. thanks the Flatiron Institute for its hospitality during his time as a long-time visitor. The Flatiron Institute is a division of the Simons Foundation.

\bibliography{literatur.bib}

\begin{thebibliography}{73}
\bibitem [{fn5()}]{fn5}%
  \BibitemOpen
  \href@noop {} {}\bibinfo {note} {To allow for a comparison on an equal
  footing, we only switch between TRIQS/CTHYB~\cite{TRIQS/CTHYB} and our MPS
  solver, while performing the DMFT loop in the same way using the TRIQS
  library~\cite{TRIQS} and the TRIQS/DFTTools
  package~\cite{TRIQS/DFTTools}.}\BibitemShut {Stop}%
\bibitem [{fn6()}]{fn6}%
  \BibitemOpen
  \href@noop {} {}\bibinfo {note} {The computational cost is rising
  substantially with $L$ due to increasing entanglement present in the
  impurity-bath system, but this is lower than the exponential cost involved in
  exact diagonalization approaches. Hence, further improvements in code
  efficiency should make larger baths accessible.}\BibitemShut {Stop}%
\end{thebibliography}

\clearpage
\setcounter{page}{0}
\thispagestyle{empty}
\onecolumngrid

\begin{center}
	\vspace{0.1cm}
	{\bfseries\large Supplemental Material for \\
		``Imaginary-time matrix product state impurity solver in a real material
		calculation: Spin-orbit coupling in Sr$_2$RuO$_4$''}\\
	\vspace{0.4cm}
	Nils-Oliver Linden,$^1$
	Manuel Zingl,$^2$
	Claudius Hubig,$^3$
	Olivier Parcollet,$^{2,4}$
	Ulrich Schollw\"ock$^{1,5}$
	\\
	\vspace{0.1cm}
	{\it\normalsize
		$^1$Arnold Sommerfeld Center of Theoretical Physics, Department of Physics,\\ University of Munich, Theresienstrasse 37, 80333 Munich, Germany\\
		$^2$Center for Computational Quantum Physics, Flatiron Institute, 162 5th Avenue, New York, NY 10010, USA \\
		$^3$Max-Planck-Institute for Quantum Optics, Hans-Kopfermann-Strasse 1, 85748 Garching, Germany \\
		$^4$Institut de Physique Th\'eorique (IPhT), CEA, CNRS, UMR 3681, 91191 Gif-sur-Yvette, France\\
		$^5$Munich Center for Quantum Science and Technology (MCQST), Schellingstrasse 4, 80799 Munich, Germany
	} \\
	(Dated: \today)
	\vspace{0.3cm}
\end{center}
\vspace{1.0cm}
\twocolumngrid

In this Supplemental Material, we provide algorithms details on the ground-state search, imaginary-time evolution, fictitious temperature, and the bath size. Additionally, we show the self-energy matrix for the calculation with SOC included.

\section{Ground-state search}
The correlated pair hopping contained in the Hubbard-Kanamori Hamiltonian~$\hat{H}_{\text{int}}$ changes the number of electrons in each impurity orbital plus its bath in steps of 2, decoupling the Hilbert space into eight sectors for an odd or even number of electrons in each of the three orbitals. In such a case, the standard MPS practice of ensuring ground-state convergence into the sector containing the global minimum is to add weak coupling terms that are reduced to zero during sweeping. We found this to be unreliable. With hindsight, we also found this procedure to leave traces in the wave function leading to higher entanglement and a less efficient encoding, even where it ends up in the correct sector at the right energy. This can be circumvented by implementing three
additional quantum numbers (beyond $SU(2)$ symmetry and
global particle number) measuring the parity in each of the sectors, allowing us to select any specific sector from the start.
Concurrent ground-state searches in each of those sectors are then
efficient and stable, and a comparison of the resulting energies reliably leads
to the global ground state.

\section{Imaginary-time evolution}

\begin{figure}[bh]
	\includegraphics[width=1.0\linewidth]{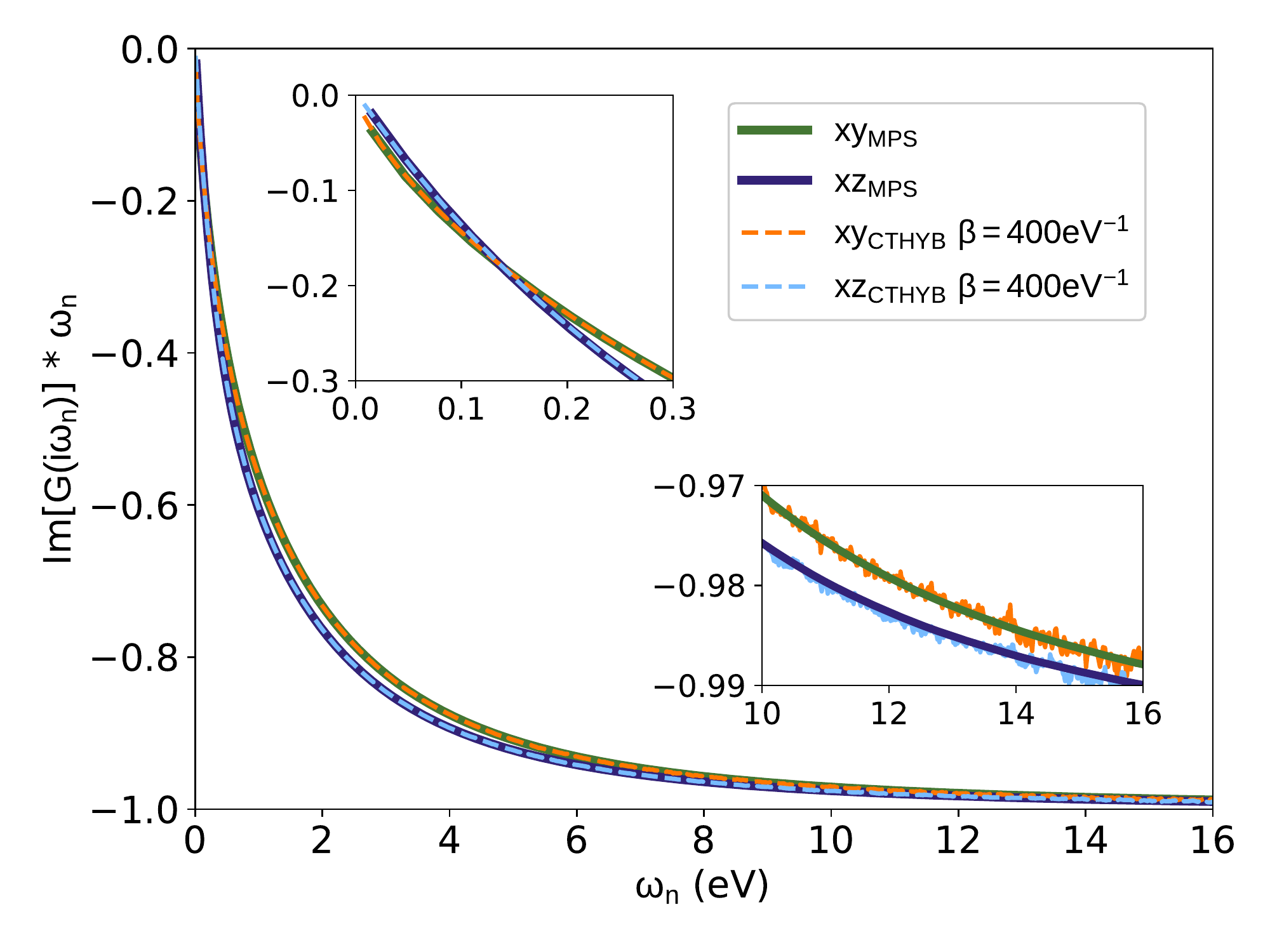}
	\caption{Im$\left[G(i\omega_n)\right] \cdot \omega_n$ for the $xy$ and $xz/yz$ orbitals
		without SOC using the MPS impurity solver and a bath size of $L=8$ (solid lines) compared to CTHYB (dashed lines) at $\beta = \SI{400}{eV^{-1}}$ (\SI{29}{K}).}
	\label{fig:figA1}
\end{figure}

For calculating Green's functions of the type $G(\tau)\equiv \langle 0|\mathcal{T}d(\tau)d^\dagger(0) | 0\rangle=\langle 0| \eul^{-\tau H} d \eul^{-\tau H} d^\dagger |0\rangle (\tau>0)$, we used the time-dependent variational principle (TDVP)~\cite{haegeman16} for a combination of high accuracy and speed for times up to $\tau \approx \SI{100}{eV^{-1}}$. As linear prediction is optimal for time series of superpositions of decaying exponentials, it allows high-precision extrapolations up to $\tau\approx \SI{2000}{eV^{-1}}$, where all $G(\tau)$ encountered were zero for practical purposes, allowing for highly precise results for $G(i\omega)$ for $\omega_n\rightarrow 0$. In the $\omega_n\rightarrow\infty$ limit, $G(i\omega) \rightarrow 1/i\omega$, which reflects the unit jump of $G(\tau)$ at $\tau=0$ due to the fermionic anticommutator $[d,d^\dagger]_+=1$. This short-time behavior is not properly captured by time-evolution methods which calculate $G(\tau)$ at finite time steps spaced by, say, $\Delta\tau=\SI{0.05}{eV^{-1}}$. Using very small timesteps for $\tau\rightarrow 0$ with these methods is not just costly, but impossible: the required state truncation/projections effectively lead back to the initial state, freezing the time evolution in a kind of `Zeno effect'. A way out is provided by Krylov-based time evolution, which uses finite timesteps but also allows the evaluation of $G(\tau)$ at all intermediate times with negligible cost, with at least the accuracy achieved for the finite timestep~\cite{paeckel2019}. In Fig.~\ref{fig:figA1} we show the perfect agreement for $G(i\omega)$ for small and large $i\omega$ between CTHYB at $\beta = \SI{400}{eV^{-1}} (\SI{29}{K})$ and the MPS result for $L=8$, providing validation for both methods. 

\section{Zero temperature}

We note that the MPS solver works directly at $\beta=\infty$, following the procedure outlined in Ref.~\cite{wolf15iii}. Then the Matsubara Green's functions become continuous, $G(i\omega_n)\rightarrow G(i\omega)$. Numerically, they are evaluated only at discrete frequencies which we choose to be the Matsubara frequencies at some fictitious temperature $\beta_{\text{fict}}$. However, this does not mean that the MPS-DMFT results reflect the physics at $\beta_{\text{fict}}$ instead of $\beta=\infty$. The MPS Green's functions were evaluated both at $\beta_{\text{fict}}=\SI{200}{eV^{-1}}$ and $\beta_{\text{fict}}=\SI{300}{eV^{-1}}$. No difference in capturing the underlying $\beta=\infty$ Green's functions was found.

\section{Bath size}

\begin{figure}[h]
	\includegraphics[width=1.0\linewidth]{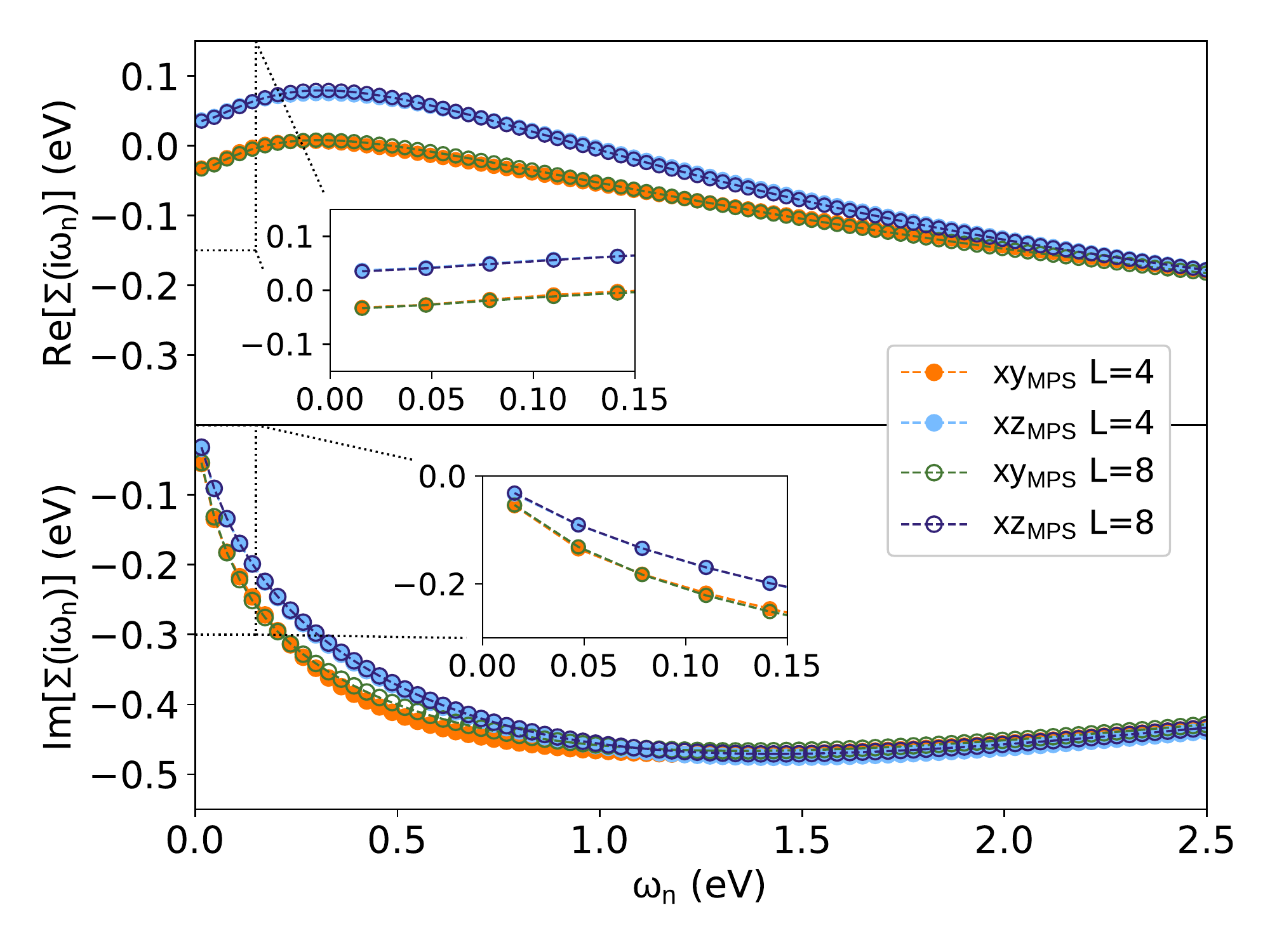}
	\caption{Comparison of the self-energies of \SRO{} without SOC for bath sizes $L=4$ and $L=8$.
		Especially at low $\omega_n$ the self-energy is already reliable for a small bath of $L=4$.}
	\label{fig:figA2}
\end{figure}

For the results without SOC presented in the main text the bath size was chosen to $L=8$ per impurity orbital and spin. Larger baths did not change results anymore, but instead `over-fitting' problems arose.
In Fig.~\ref{fig:figA2} we show that the difference between $L=8$ and $L=4$ in the self-energy is small,
especially for the lower frequencies.
Therefore, already a smaller bath size of $L=4$ provides a reasonable self-energy, justifying
the use of $L=4$ in the calculations with SOC included.

\section{Off-diagonal self-energies}
\begin{figure}[h]
	\includegraphics[width=1.0\linewidth]{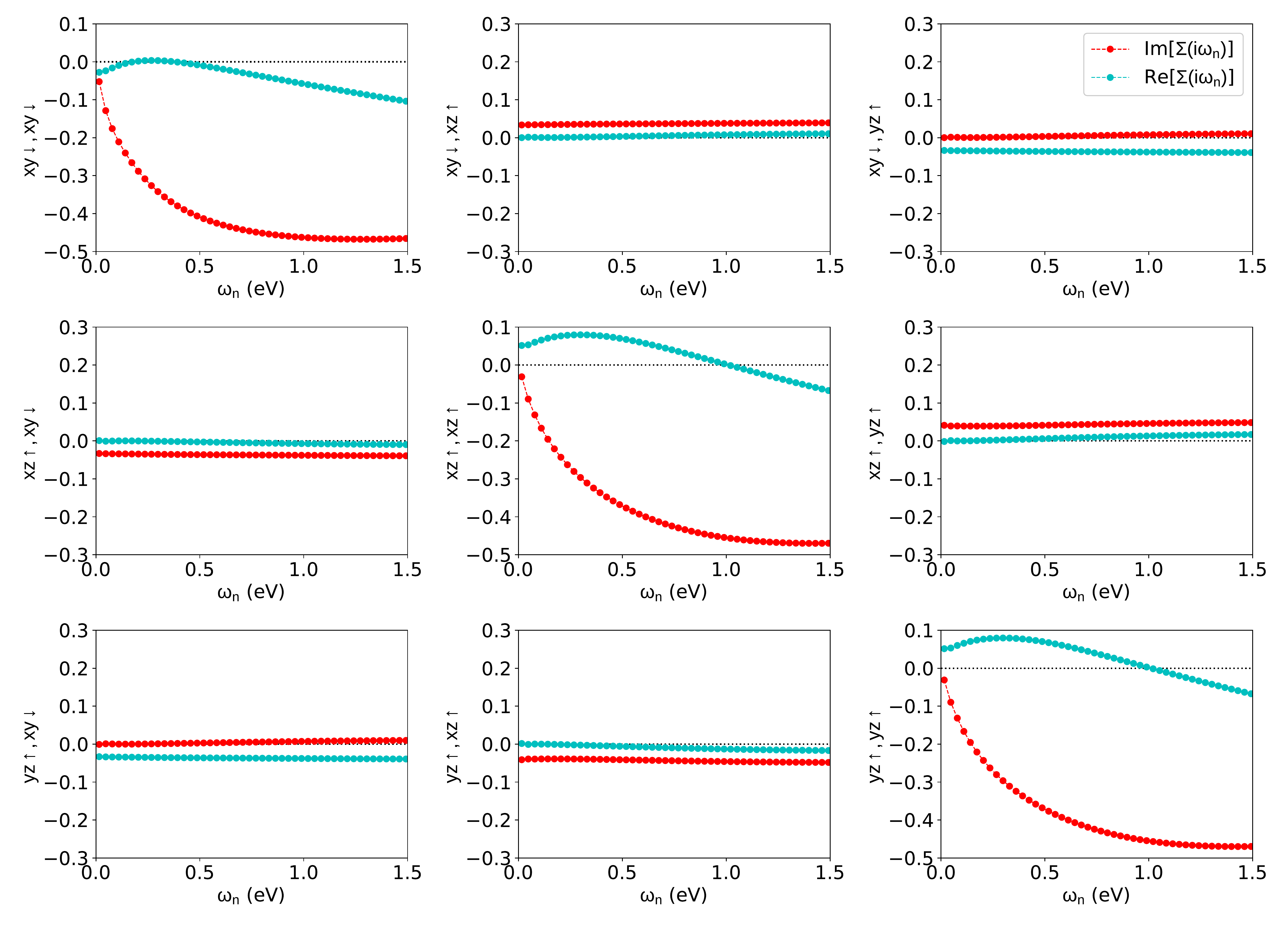}
	\caption{Self-energy matrix $\Sigma_{m\sigma,m'\sigma'}(i\omega_n)$ for the $(m\sigma,m'\sigma')$ block with $m\sigma,m'\sigma'\in \{xy\downarrow,xz\uparrow,yz\uparrow\}$ using the MPS solver for \SRO{} with SOC and a bath size of $L=4$ per orbital and spin state. For the diagonal elements the chemical potential has been subtracted from the real parts.}
	\label{fig:figA3}
\end{figure}

The SOC leads to off-diagonal elements in the $6 \times 6$ self-energy matrix $\Sigma_{m\sigma,m'\sigma'}(i\omega_n)$. However, in the
orbital basis the elements of the self-energy matrix can be grouped into two degenerate $3 \times 3$ blocks, $\{xy\downarrow,xz\uparrow,yz\uparrow\}$ and $\{xy\uparrow,xz\downarrow,yz\downarrow\}$ resp.
One block of the self-energy matrix, with $m\sigma,m'\sigma'\in \{xy\downarrow,xz\uparrow,yz\uparrow\}$, is shown in Fig.~\ref{fig:figA3}.
These results have been obtained with a bath size of $L=4$.
Note that in comparison to the diagonal elements the off-diagonal elements are almost frequency independent.

~\clearpage
\onecolumngrid

\twocolumngrid

\end{document}